\documentclass[a4paper,10pt,pre,twocolumn,showpacs,aps,floats,superscriptaddress]{revtex4}

\usepackage{epsfig}\usepackage{amsmath}

\newcommand{\avk}{\langle k \rangle}

\newcommand{\Ri}{r_{\infty}}

\newcommand{\fluck}{\langle k^2 \rangle}

\begin{document}

\title{Dynamics of Rumor Spreading in Complex Networks} 

\author{Yamir Moreno}

\affiliation{Departamento de F\'{\i}sica Te\'orica, Universidad de
Zaragoza, Zaragoza 50009, Spain}

\affiliation{Instituto de Biocomputaci\'on y F\'{\i}sica de Sistemas
Complejos, Universidad de Zaragoza, Zaragoza 50009, Spain}

\author{Maziar Nekovee}

\affiliation{Complexity Research Group, Polaris 134 BT Exact,
Martlesham, Suffolk IP5 3RE, UK}

\author{Amalio F. Pacheco} 

\affiliation{Departamento de F\'{\i}sica Te\'orica, Universidad de
Zaragoza, Zaragoza 50009, Spain}

\affiliation{Instituto de Biocomputaci\'on y F\'{\i}sica de Sistemas
Complejos, Universidad de Zaragoza, Zaragoza 50009, Spain}

\date{\today}

\widetext

\begin{abstract} 

We derive the mean-field equations characterizing the dynamics of a
rumor process that takes place on top of complex heterogeneous
networks. These equations are solved numerically by means of a
stochastic approach. First, we present analytical and
Monte Carlo calculations for homogeneous networks and compare the
results with those obtained by the numerical method. Then, we study
the spreading process in detail for random scale-free networks. The
time profiles for several quantities are numerically computed, which
allow us to distinguish among different variants of rumor spreading
algorithms. Our conclusions are directed to possible applications in
replicated database maintenance, peer to peer communication networks
and social spreading phenomena.

\end{abstract}


\maketitle

\section{Introduction}
\label{section0}

During the last years, many systems have been analyzed from the
perspective of graph theory \cite{doro,bara02}. It turns out that
seemingly diverse systems such as the Internet, the World Wide Web
(WWW), metabolic and protein interaction networks and food webs, to
mention a few examples, share many topological properties
\cite{strogatz}. Among these properties, the fact that one can go from
one node (or element) of the network to another node passing by just a
few others is perhaps the most popular property, known as ``six
degrees of separation'' or small-world (SW) property
\cite{strogatz,ws98}. The SW feature has been shown to improve the
performance of many dynamical processes as compared to regular
lattices; a direct consequence of the existence of key shortcuts that
speed up the communication between otherwise distant nodes and of the
shorter path length among any two nodes on the net
\cite{doro,bara02,strogatz}.

However, it has also been recognized that there are at least two types
of networks fulfilling the SW property but radically different as soon
as dynamical processes are ran on top of them. The first type can be
called ``exponential networks'' since the probability of finding a
node with connectivity (or degree) $k$ different from the average
connectivity $\avk$ decays exponentially fast for large $k$
\cite{ama00}. The second kind of networks comprises those referred to
as ``scale-free'' (SF) networks \cite{bar99}. For these networks, the
probability that a given node is connected to $k$ other nodes follows
a power-law of the form $P(k) \sim k^{-\gamma}$, with the remarkable
feature that $\gamma\le 3$ for most real-world networks
\cite{doro,bara02}.

The heterogeneity of the connectivity distribution in scale-free
networks greatly impacts the dynamics of processes that they
support. One of the most remarkable examples is that an epidemic
disease will pervade in an infinite-size SF network regardless of its
spreading rate \cite{pv01a,moreno02,virusreview,av03,n02b}.  The
change in the behavior of the processes is so radical in this case
that it has been claimed that the standard epidemiological framework
should be carefully revisited.  This might be bad news for
epidemiologists, and those fighting natural and computer viruses. On
the other hand, in a number of important technological and commercial
applications, it is desirable to spread the ``epidemic'' as fast and
as efficient as possible, not to prevent it from spreading. Important
examples of such applications are epidemic (or rumor-based) protocols
for data dissemination and resource discovery on the Internet
\cite{vogels,ep3,p2p,ep1}, and marketing campaigns using rumor-like
strategies (viral marketing).

The above applications, and their dynamics, have passed almost
unnoticed \cite{zanette,liu} to the physics community working on
complex networks despite the fact that they have been extensively
studied by computer scientists and sociologists \cite{ep1,dkbook}. The
problem here consists of designing an epidemic (or rumor-mongering)
algorithm in such a way that the dissemination of data or information
from any node of a network reaches the largest possible number of
remaining nodes.  Note that in this case, in contrast to epidemic
modeling, one is free to design the rules of epidemic infection in
order to reach the desired result, instead of having to model an
existing process. Furthermore, in a number of applications, such as
peer-to-peer file sharing systems \cite{p2p} built on top of the
Internet and grid computing \cite{grid}, the connectivity distribution
of the nodes can also be changed in order to maximize the performance
of such protocols.

In this paper we study in detail the dynamics of a generic rumor model
\cite{dk64} on complex scale-free topologies through analytic and
numerical studies, and investigate the impact of the interaction rules
on the efficiency and reliability of the rumor process. We first solve
the model analytically for the case of exponential networks in the
infinite time limit and then introduce a stochastic approach to deal
with the numerical solution of the mean-field rate equations
characterizing the system's dynamics. The method
\cite{pre98,jgr,mgp03} is used to obtain accurate results for several
quantities when the topology of random SF networks is taken into
account, without using large and expensive Monte Carlo (MC)
simulations. The rest of the paper is organized as follows. Section II
is devoted to introducing the rumor model and to derive the mean-field
rate equations used throughout the paper. In Section III we deal with
the stochastic approach, and compare its performance with analytical
and MC calculations in homogeneous systems. We extend the method to
the case of power-law distributed networks and present the results
obtained for this kind of networks in Section IV and V. Finally, the
paper is rounded off in the last Section, where conclusions are given.

\section{Rumor Model in Homogeneous Networks}
\label{section1}

The rumor model is defined as follows. Each of the $N$ elements of the
network can be in three different states. Following the original
terminology and the epidemiological literature \cite{dkbook}, these
three classes correspond to ignorant, spreader and stifler
nodes. Ignorants are those individuals who have not heard the rumor
and hence they are susceptible to be informed. The second class
comprises active individuals that are spreading the rumor. Finally,
stiflers are those who know the rumor but that are no longer spreading
it. The spreading process evolves by directed contacts of the
spreaders with others in the population. When a spreader meets an
ignorant the last one turns into a new spreader with probability
$\lambda$. The decay of the spreading process may be due to a
mechanism of ``forgetting'' or because spreaders learn that the rumor
has lost its ``news value''. We assume this latter hypothesis as the
most plausible so that the contacting spreaders become stiflers with
probability $\alpha$ if they encounter another spreader or a
stifler. Note that as we are designing our rumor strategy in such
a way that the fraction of the population which ultimately learns the
rumor be the maximum possible, we have assumed that contacts of the
type spreader-spreader are directed, that is, only the contacting
individual loses the interest in propagating the rumor
further. Therefore, there is no double transition to the stifler
class.

In a homogeneous system, the original rumor model due to Daley and
Kendall \cite{dk64} can be described in terms of
the densities of ignorants, spreaders, and stiflers, $i(t)$, $s(t)$,
and $r(t)$, respectively, as a function of time. Besides, we have the
normalization condition,
\begin{equation}
i(t)+s(t)+r(t)=1. \label{eq1}
\end{equation}
In order to obtain an analytical insight and a way to later test our
numerical approach, we first study the rumor model on top of
exponentially distributed networks. These include models of random
graphs as well as the Watts and Strogatz (WS) small-world model
\cite{ws98,strogatz}. This model produces a network made up of $N$
nodes with at least $m$ links to other nodes. The resulting
connectivity distribution in the random graph limit of the model
\cite{ws98} takes the form
%\begin{equation}
\[
P(k)=\frac{m^{k-m}}{(k-m)!}e^{-m},
\]
%\end{equation}
which gives an average connectivity $\avk=2m$. Hence, the probability
that a node has a degree $k\gg \avk$ decays exponentially fast and the
network can be regarded as homogeneous.

The mean-field rate equations for the evolution of the three densities
satisfy the following set of coupled differential equations:
\begin{eqnarray}
\frac{d i(t)}{d t} & = & - \lambda \avk i(t) s(t), \label{eq2}\\
\frac{d s(t)}{d t} & = & \lambda \avk i(t) s(t) - \alpha \avk s(t)
[s(t)+r(t)], \label{eq3}\\ 
\frac{d r(t)}{d t} & = & \alpha \avk s(t) [s(t)+r(t)], \label{eq4}
\end{eqnarray}
with the initial conditions $i(0)=(N-1)/N$, $s(0)=1/N$ and
$r(0)=0$. The above equations state that the density of spreaders
increases at a rate proportional to the spreading rate $\lambda$, the
average number of contacts of each individual $\avk$ and to the
densities of ignorant and spreader individuals, $i(t)$ and $s(t)$,
respectively. On the other hand, the annihilation mechanism considers
that spreaders decay into the stifler class at a rate $\alpha \avk$
times the density of spreaders and of non-ignorant individuals
$1-i(t)=s(t)+r(t)$.

The system of differential equations\ (\ref{eq2}-\ref{eq4}) can be
analytically solved in the infinite time limit when
$s(\infty)=0$. Using equation\ (\ref{eq1}), we have that
$\int_{0}^{\infty} s(t) dt=r_{\infty}=lim_{t\rightarrow\infty}r(t)
$. Introducing the new variable $\beta=1+\lambda/\alpha$ we obtain the
transcendental equation,
\begin{equation}
r_{\infty}=1-e^{-\beta r_{\infty}}. \label{eq5}
\end{equation}

Equation\ (\ref{eq5}) always admits the trivial solution
$r_{\infty}=0$, but at the same time it also has another physically
relevant solution {\em for all} values of the parameters
$\lambda$ and $\alpha$. This can be easily appreciated since the condition,
\begin{equation}
\frac{d}{dr_{\infty}}\left. \left(1-e^{-\beta r_{\infty}}\right )
\right |_{\Ri=0}>1,
\end{equation}
reduces to $\lambda/\alpha > 0$. That is, there is no ``rumor
threshold'' contrary to the case of epidemic spreading
\cite{moreno02}. This strikingly different behavior does not come from
any difference in the growth mechanism of $s(t)$ $-$the two are
actually the same$-$, but from the disparate rules for the decay of
the spreading process.

On the other hand, this result also points out that a mathematical
model for the spreading of rumors can be constructed in many different
ways. The results of this paper, however, indicate that the presence
of spreader annihilation terms due to spreader-spreader and
spreader-stifler interactions is very relevant for practical
implementations \cite{mnv03,nm03}. We shall come back to this point
later on.

\section{Stochastic Numerical Approach}
\label{section2}

Recently \cite{mgp03}, we have introduced a numerical technique
\cite{pre98} to deal with the mean-field rate equations appearing in
epidemic-like models. It solves the differential equations by
calculating the passage probabilities for the different
transitions. The main advantage of this method, as compared to MC
simulations, is its modest memory and CPU time requirements for large
system sizes. Besides, we do not have to generate any
network. Instead, we produce a sequence of integers distributed
according to the desired connectivity distribution $P(k)$. The
numerical procedure here proceeds as follows. At each time step until
the end of the rumor spreading process, the following steps are
performed:

\begin{table}[t]
\begin{ruledtabular}
\begin{tabular}{|c|c|c|c|}
\hline
$\alpha$ & Eq.\ (\ref{eq5}) & MC & SNA \\
\hline
1 & 0.7968 & 0.813 & 0.802 \\
0.5 & 0.9404 & 0.962 & 0.954 \\
0.25 & 0.9930 & 0.986 & 0.987 \\
0.2 & 0.9974 & 0.996 & 0.997 \\
0.1 & 0.9999 & 0.998 & 0.999 \\
\hline
\hline
\end{tabular}
\caption{Density of stiflers at the end of the rumor spreading
process. Results are shown for 5 different values of $\alpha$ for each
method considered. Monte Carlo (MC) simulations were performed in a WS
network with $\avk=6$ and $N=10^4$ nodes. The same system size was
used in the stochastic numerical approach (SNA).}
\label{table1}
\end{ruledtabular}
\end{table}

\begin{enumerate}
\item Identify from the mean-field rate equations the transition
probabilities per time unit from one state into the following one,
that is, from the $i$ class to the $s$ class, $W_{i\rightarrow s}$,
and finally to the $r$ class, $W_{s\rightarrow r}$.
\item Calculate the mean time interval, $\tau$, for one transition to
occur. This is determined as the inverse of the sum of all the
transition probabilities; $\tau=1/(W_{i\rightarrow s}+W_{s\rightarrow
r})$.
\item Stochastically decide what transition will actually take
place. This is done by deciding that the probabilities for both
transitions are given by $\Pi_{i\rightarrow s}=W_{i\rightarrow s}\tau$
and $\Pi_{s\rightarrow r}=W_{s\rightarrow r}\tau$, respectively,
materializing the choice by generating a random number between 0 and
1.
\end{enumerate}

The numerical algorithm described above does not depend on the
topological features of the network on top of which the rumor dynamics
is taking place. Indeed, all the topological information, including
correlations, enters in the computation of the transition
probabilities. We should note here that the present results are
obtained for uncorrelated networks. The method could also be applied
to correlated networks without explicit generation of them. In that
case, one should work with the two point correlation function
$P(k,k')$ \cite{mgp03} instead of using $P(k)$. On the other hand, a
correlated network could be built up as in \cite{vw}.

In order to gain confidence with the method and to show its soundness,
we show in Table\ \ref{table1} the values of $r_{\infty}$ obtained
from Eq.\ (\ref{eq5}), MC simulations and the stochastic approach for
homogeneous networks. In this case, the transition probabilities are
the same for all the elements within a given class ($i$, $s$ or $r$)
irrespective of their actual connectivities. From equations\
(\ref{eq2}-\ref{eq4}) we get
\begin{eqnarray}
W_{i\rightarrow s}(t) & = & N \lambda\avk i(t)s(t), \\
W_{s\rightarrow r}(t) & = & N \alpha\avk s(t)[s(t)+r(t)],
\end{eqnarray}
for the transitions from the ignorant to the spreader class and from
the spreader to the stifler class, respectively. 

It can be seen from Table\ \ref{table1} that the difference between
the SNA result and the MC simulations is less that $1.4\%$, indicating
the reliability of the SNA approach.  The remaining small differences
between the SNA and the MC results is mainly due to the fact that the
homogeneous SNA model does not take into account the exponentially
decaying fluctations in the connectivity of WS networks. On the other
hand, MC simulations of the rumor dynamics for a network made up of
$N=10^4$ nodes, averaged over at least 10 different network
realizations and 1000 iterations, took several hours. Eventually, this
method takes up to a few days when increasing the system size and
decreasing the value of $\alpha$. On the contrary, the stochastic
approach is very fast. Indeed, for the same parameter values, the
numerical simulation takes around 5 minutes CPU time in a 2.0Ghz-P4
PC. Therefore, having such a method will allow us to scrutinize very
efficiently and accurately the whole phase diagram and time profiles
of the process under study.  In what follows, we analyze in detail the
dynamics of the rumor spreading process by numerically solving the
mean-field rate equations for SF networks.

\section{power-law distributed networks}

The heterogeneity of the connectivity distribution inherent to SF
networks significantly affects the dynamical evolution of processes
that take place on top of these networks
\cite{pv01a,moreno02,virusreview,av03,n02b,havlin01,newman00,bar00,mv02}. We
have learned in recent years that the fluctuations of the connectivity
distribution, $\fluck$, can not be neglected even for finite size
systems \cite{virusreview}. Thus, the system of differential
equations\ (\ref{eq2}-\ref{eq4}) should be modified accordingly. In
particular, we should take into account that nodes could not only be
in three different states, but also they belong to different
connectivity classes $k$. Let us denote by $i_k(t)$, $s_k(t)$, and
$r_k(t)$ the densities of ignorants, spreaders and stiflers with
connectivity $k$, respectively. In addition, we have that
$i_k(t)+s_k(t)+r_k(t)=1$. The mean-field rate equations now read as,
\begin{eqnarray}
\frac{d i_k(t)}{d t} & = & - \lambda k i_k(t)\sum_{k'} \frac{k'
P(k')s_{k'}(t)}{\avk}, \label{eq9}\\
\frac{d s_k(t)}{d t} & = & \lambda k i_k(t)\sum_{k'} \frac{k'
P(k')s_{k'}(t)}{\avk}\nonumber\\
&  & -\alpha k s_k(t) \sum_{k'} \frac{k'
P(k')[s_{k'}(t)+r_{k'}(t)]}{\avk} , \label{eq10}\\
\frac{d r_k(t)}{d t} & = & \alpha k s_k(t) \sum_{k'} \frac{k'
P(k')[s_{k'}(t)+r_{k'}(t)]}{\avk} , \label{eq11}
\end{eqnarray}
where $P(k)$ is the connectivity distribution of the nodes and
$\sum_{k'} k'P(k')s_{k'}(t)/\avk$ is the probability that any given
node points to a spreader. We start from a randomly selected spreader
and all the remaining nodes in the ignorant class. The summation in
Eq.\ (\ref{eq10}) stands for the probability that a node points to a
spreader or a stifler. Note that as before, we do not allow for double
transitions from the spreader to the stifler class. Next, we compute
the respective transition probabilities. In this case, we should also
consider that transitions from one state into another also take place
within connectivity classes. Thus, the transition probabilities depend
on $k$ as well. From Eq.\ (\ref{eq10}) we obtain,
\begin{eqnarray}
W_{i\rightarrow s}(t,k) & = & \lambda k N P(k) i_k(t)
\sum_{k'}\frac{k'P(k')s_{k'}(t)}{\avk},
\label{eq12}\\
W_{s\rightarrow r}(t,k) & = & \alpha k N P(k) s_k(t) \sum_{k'}\frac{k'
P(k')[s_{k'}(t)+r_{k'}(t)]}{\avk}, \nonumber\\
& & \label{eq13}
\end{eqnarray}
where all the topological information is contained. Finally for the
mean time interval after $i-1$ transitions, $\tau$, we find at each
time step
\begin{equation}
\tau=\frac{1}{W_{i\rightarrow s}(t)+W_{s\rightarrow r}(t)}, \label{eq14}
\end{equation}
with $W_{i\rightarrow s}(t)=\sum_k W_{i\rightarrow s}(t,k)$,
$W_{s\rightarrow r}(t)=\sum_k W_{s\rightarrow r}(t,k)$ and
$t=\sum_j^{i-1}\tau_j$, where the $\tau_j$s are the mean times of the
$i-1$ previous transitions. At this point, the identification of what
transition takes place and which connectivity class is affected
proceeds as defined in step $3$ of the previous section.

\section{results and discussion}

\begin{figure}
\begin{center}
\epsfig{file=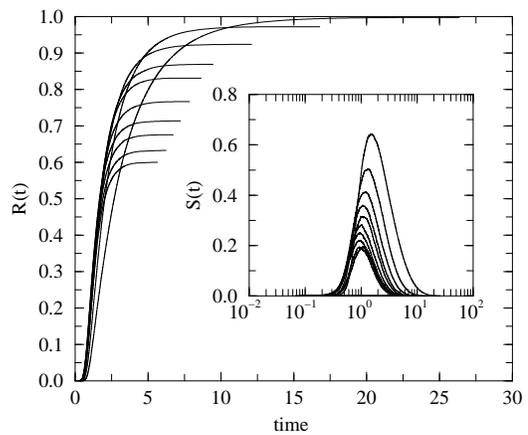,width=2.3in,angle=-90,clip=1}
\end{center}
\caption{Time evolution of the density of stifler individuals for
  different values of $\alpha$. From below, the values of $\alpha$ go
  from $1.0$ to $0.1$ at fixed increments of $0.1$. The inset shows
  the time dependency of the density of spreaders. The system size is
  $N=10^4$, $\avk=6$ and $\gamma=3$. Time is in units of
  $\alpha^{-1}$.}
\label{figure1}
\end{figure}
The stochastic method described above can be used to explore several
quantities characterizing the dynamics of the rumor spreading
process. Throughout the rest of the paper we set $\lambda=1$ without
loss of generality and vary the value of $\alpha$. We first generated
a sequence of integers distributed according to $P(k)\sim k^{-\gamma}$
with $\gamma=3$ and $\avk=6$. As initial condition we use
$r_k(t=0)=0$, and
\begin{equation}
s_k(t)=
\begin{cases}
\frac{1}{NP(k)} 
& k=k_{i}
\\
0          & \text{otherwise}
\end{cases}
\end{equation}
where $k_i$ is the connectivity of the randomly chosen initial
spreader. The results are then averaged over at least $1000$ different
choices of $k_i$.

One of the most important practical aspects of any rumor mongering
process is whether or not it reaches a high number of
individuals. This magnitude is simply given by the final density of
stiflers and is called {\em reliability} of the rumor
process. However, it is also of great importance for potential
applications that higher levels of reliability are reached as fast as
possible, which constitute a practical measure of the cost associated
to such levels of stiflers. For example, in technological
applications, where one may consider several strategies
\cite{mnv03,nm03}, it is possible to define a key global quantity, the
efficiency of the process, which is the ratio between the reliability
and the traffic imposed to the network. For these applications it is
not only important to have high levels of reliability but also to
achieve these with the lowest possible load resulting from the
epidemic protocol's message passing traffic. This is important in
order to avoid network congestion and also to reduce the amount of
processing power used by nodes participating in the rumor process.

\begin{figure}[t]
\begin{center}
\epsfig{file=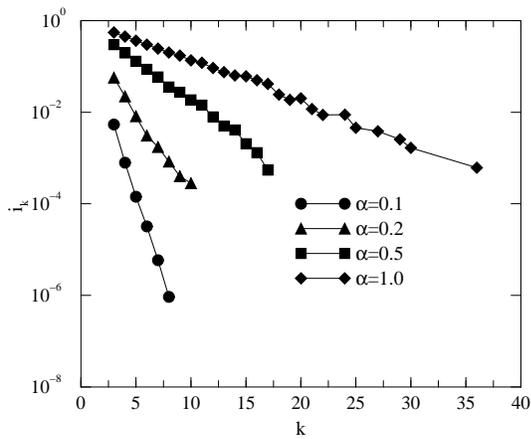,width=2.3in,angle=-90,clip=1}
\end{center}
\caption{Density of ignorants $i_k$ at the end of the rumor process as
  a function of their connectivity $k$. A clear exponential decay can
  be appreciated for all values of $\alpha$ shown. This implies that
  hubs have efficiently learned the rumor.}
\label{figure2}
\end{figure}

In order to analyze, from a global perspective, this trade-off between
reliability and cost, we use time as a practical measure of
efficiency. We call a rumor process less efficient than another if it
needs more time to reach the same level of reliability. Figure\
\ref{figure1} shows the time evolution of the density of stiflers for
several values of the parameter $\alpha$. It turns out, as expected,
that the number of individuals who finally learned the rumor increases
as the probability of becoming stifler decreases. On the other hand,
the time it takes for $R(t)$ to reach its asymptotic value slightly
increases with $\alpha^{-1}$, but clear differences do not arise for
the two extreme values of $\alpha$. In fact, for a given time after
the beginning of the rumor propagation, the density of stiflers scales
with the inverse of $\alpha$. This behavior is further corroborated in
the inset, where the growth of the density of spreaders as time goes
on is shown for the same values of the parameter $\alpha$. While the
peaks of the curves get larger and larger, the times at which the
maxima are reached are of the same order of magnitude and thus the
meantimes of the spreading processes do not differ significantly.

\begin{figure}[t]
\begin{center}
\epsfig{file=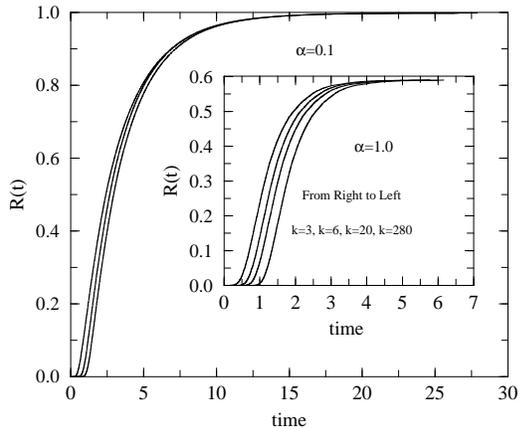,width=2.3in,angle=-90,clip=1}
\end{center}
\caption{Density of stiflers as a function of time for $\alpha=0.1$
  (main figure) and $\alpha=1.0$ (inset) when the initial spreader has
  the connectivity indicated in the inset. Note that in all cases the
  final density of individuals who have learned the rumor is the same,
  but the asymptotic value is reached at different times. The model
  parameters are as of Fig.\ (\ref{figure1}). Time is in units of
  $\alpha^{-1}$.}
\label{figure3}
\end{figure}

Figure\ \ref{figure2} shows another aspect worth taking into account
when dealing with rumor algorithms. For a given level of reliability,
it is also of interest to know the distribution of ignorants (or
stiflers) by classes $k$. The figure shows a coarse-grained picture of
Fig.\ \ref{figure1}, where the density of ignorants $i_k$ according to
the connectivity of the individuals has been represented for different
values of $\alpha$. The results indicate that the probability of
having an ignorant with a connectivity $k$, at the end of the rumor
propagation, decays exponentially fast with a sharp cut-off $k_c$ for
large connectivity values, which depends on $\alpha$. In fact, $k_c$ is
always well below the natural cut-off of the connectivity distribution
($\sim 10^2$) even for small values of $\alpha$. This implies that
hubs effectively learn the rumor.

We can further scrutinize the dynamics of the rumor spreading process
by looking at the final density of stiflers when the initial spreader
has a given connectivity $k_i$. Figure\ \ref{figure3} represents the
reliability as a function of time (in units of $\alpha^{-1}$) when the
rumor starts propagating from a node of connectivity $k_i=k_{min}=3$,
$k_i=\avk=6$, $k_i=20$ and $k_i=k_{max}\sim 280$ for two different
values of $\alpha$: $0.1$ (main figure) and $1.0$
(inset). Interestingly, the final value of $R(t)$ does not depend on
the initial seed, but reaches the same level irrespective of the
connectivity of the very first spreader $k_i$. This is a genuine
behavior of the rumor dynamics and is the opposite to what has been
observed in other epidemic models like the SIR model \cite{moreno02},
where the final number of recovered individuals strongly depends on
the connectivity of the initially infected individuals. However, a
closer look to the spreading dynamics tell us that not all is the same
for different initial spreaders.

The figure also indicates that as the connectivity of the seed is
increased, the time it takes for the rumor to reach the asymptotic
value decreases, so that for a fixed time length the number of
individuals in the stifler class is higher when $k_i$ gets
larger. This feature suggests an interesting alternative for practical
applications: start propagating the rumor from the most connected
nodes. Even in case that no direct link exits between a node that is
willing to spread an update and a hub, a dynamical (or temporal)
shortcut to a well connected node could be created in order to speed
up the process. With this procedure, the density of stiflers at the
intermediate stages of the spreading process could be as much
different as $30\%$ for moderate values of $\alpha$. This translates
in less costs, because one can always implement an algorithm that kill
off the actual spreading when a given level of reliability is
reached. Note, however, that this behavior slightly depends on
$\alpha$, being the differences always appreciable, but more important
as $\alpha$ increases.

Finally, we have exploited the fastness of the stochastic approach
used here to explore the consequences of implementing three different
annihilation rules for the rumor spreading decay. In particular, we
consider that the spreading process dies out proportionally only to
the number of spreaders ($ss$ interactions) or to the number of
stiflers ($sr$ interactions). This modifies the terms entering in the
sum of Eqs.\ (\ref{eq10}-\ref{eq11}) so that now the transition
probabilities from the $s$ into the $r$ class read

\begin{eqnarray}
W_{s\rightarrow r}^{ss}(t,k) & = & \alpha k N P(k) s_k(t)
\sum_{k'}\frac{k'P(k')s_{k'}(t)}{\avk},
\label{eq15}\\
W_{s\rightarrow r}^{sr}(t,k) & = & \alpha k N P(k) s_k(t)
\sum_{k'}\frac{k' P(k')r_{k'}(t)}{\avk},
\label{eq16}
\end{eqnarray}
respectively. Table\ \ref{table2} summarizes the reliability of the
process as a function of $\alpha$ for the three mechanisms considered
\cite{note1}. The results indicate that in all variants, the final
density of stifler individuals is higher than for the ``classical''
setting. However, in order to evaluate the efficiency of the process
from a global perspective, we must look at the time evolution of the
densities as we did before.

\begin{table}[t]
\begin{ruledtabular}
\begin{tabular}{|c|c|c|c|}
\hline
$\alpha$ & $R^{s(s+r)}$ & $R^{sr}$  & $R^{ss}$ \\
\hline
1 & 0.592 & 0.857 & 0.985 \\
0.9 & 0.635 & 0.886 & 0.989 \\
0.8 & 0.674 & 0.911 & 0.991 \\
0.7 & 0.710 & 0.938 & 0.993 \\
0.6 & 0.766 & 0.960 & 0.993 \\
0.5 & 0.818 & 0.967 & 0.997 \\
0.4 & 0.871 & 0.980 & 0.997 \\
0.3 & 0.925 & 0.997 & 0.998 \\
0.2 & 0.962 & 0.999 & 0.999 \\
0.1 & 0.988 & 0.999 & 0.999 \\
\hline
\hline
\end{tabular}
\caption{Density of stiflers at the end of the rumor spreading
process. Results are shown for 10 different values of $\alpha$ for each
annihilation term considered. Simulations were performed for a network
with $\avk=6$ and $N=10^4$ nodes. See the text for further details.}
\label{table2}
\end{ruledtabular}
\end{table}

\begin{figure}[t]
\begin{center}
\epsfig{file=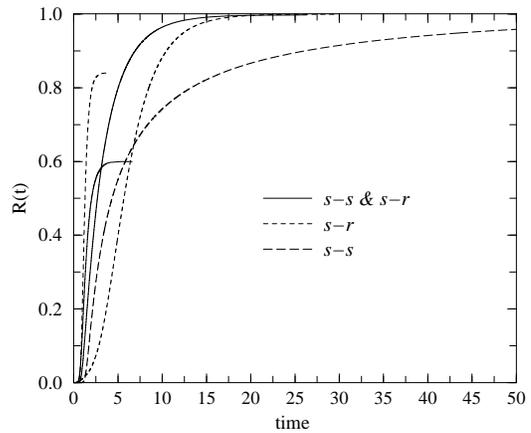,width=2.3in,angle=-90,clip=1}
\end{center}
\caption{Stifler's growth as a function of time for three different
  annihilation mechanisms as explained in the text. Curves show the
  time profiles for the two extreme values of $\alpha$ used in the
  simulations: $\alpha=1.0$ (lower curves) and $0.1$ (higher
  curves). The curve for $s-s$ interactions is for $\alpha=0.1$ and is
  not complete for clarity. Time is in units of $\alpha^{-1}$.}
\label{figure4}
\end{figure}

\begin{figure}[t]
\begin{center}
\epsfig{file=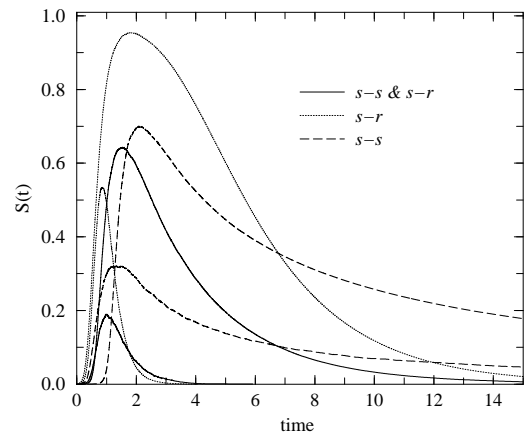,width=2.3in,angle=-90,clip=1}
\end{center}
\caption{Growth and decay of the populations of spreaders when the
  annihilation mechanism includes interactions of the type $s-s$ and
  $s-r$ or only $s-r$ or $s-s$. Curves show the time profiles for the
  two extreme values of $\alpha$ used in the simulations: $\alpha=1.0$
  (lower curves) and $0.1$ (higher curves). Note that although the
  final number of stiflers when only $s-s$ interactions enter in the
  decay mechanism, the time it takes for the rumor to reach the
  asymptotic value is very high as compared to the other two
  mechanisms. Curves corresponding to the $s-s$ interactions are not
  complete for clarity. Time is in units of $\alpha^{-1}$.}
\label{figure5}
\end{figure}

In Figs.\ \ref{figure4} and\ \ref{figure5} we have represented the
time (in units of $\alpha^{-1}$) profiles of $R(t)$ and $S(t)$ for
each decay term and several values of $\alpha$. From the figures, it
is clear that while the final density of stiflers increases when
modifying the original decay rules, the time needed to reach such high
levels of reliability also increases. This is due to the fact that the
tails of the densities of spreaders decay more slowly than before. In
particular, it is noticeable that when only spreader-spreader
interactions are taken into account in the decay mechanism, the
lifetime of the propagation process is more than two times longer than
for the other two settings. This means that this implementation is not
very suitable for practical applications as the costs associated to
the process rise as well. On the other hand, the performance of the
spreader-stifler setting seems to depend on the value of $\alpha$ in
such a way that it is more efficient at both the reliability level and
time consumption for a large $\alpha$, but not in the middle region of
the parameter space. In summary, the present results support that the
original model works quite well under any condition, while other
variants can be considered depending on the value of $\alpha$ used and
the type of applications they are designed for.

\section{conclusions}

In this paper, we have analyzed the spreading dynamics of rumor models
in complex heterogeneous networks. We have first introduced a useful
stochastic method that allows us to obtain meaningful time profiles
for the quantities characterizing the propagation process. The method
is based on the numerical solution of the mean-field rate equations
describing the model, and contrary to Monte Carlo simulations, there
is no need of generating explicitly the network. This allows to save
memory and a fast exploration of the whole evolution diagram of the
process.

The kind of processes studied here are of great practical importance
since epidemic data dissemination might become the standard practice
in multiple technological applications. The results show that there is
a fragile balance between different levels of reliability and the
costs (in terms of time) associated to them. In this sense, our study
may open new paths in the use of rumor-mongering process for
replicated database maintenance, reliable group communication and peer
to peer networks \cite{dave,deering,vogels,ep3,p2p,ep1}. Besides, as
shown here, the behavior and features of the different algorithms one
may implement are not trivial and depend on the type of mechanisms
used for both the creation and the annihilation terms. It is worth
noting here that we have studied the simplest possible set of rumor
algorithms, but other ingredients such as memory must be incorporated
in more elaborated models \cite{mnv03,nm03}.

Of further interest would be a more careful exploration of the
possibility of using dynamical shortcuts for a more efficient
spreading of the updates. Our results suggest that it would be more
economic to start from hubs and then kill off the updating process
when a given level of reliability is reached than starting at random
and letting the process dies out by itself. Preliminary studies of
more elaborated models aimed at implementing a practical protocol
confirm our results \cite{nm03}. This feature is specially relevant
for the understanding and modeling of social phenomena such as the
spreading of new ideas or the design of efficient marketing campaigns.

\begin{acknowledgments}
We would like to thank A.\ Vespignani for many useful discussions and
comments. Y.\ M.\ thanks M.\ V\'azquez-Prada for helpful comments and
the hospitality of BT Exact, U.K. where parts of this work were
carried out. Y.\ M.\ acknowledges financial support from the
Secretar\'{\i}a de Estado de Educaci\'on y Universidades (Spain,
SB2000-0357). This work has been partially supported by the Spanish
DGICYT project BFM2002-01798.
\end{acknowledgments}


\begin{thebibliography}{99}

\bibitem{doro} S. N. Dorogovtsev and J. F. F. Mendes, Adv. Phys. {\bf
51}, 1079 (2002).

\bibitem{bara02} R. Albert and A.-L. Barab\'{a}si,
Rev. Mod. Phys. {\bf 74}, 47 (2002).

\bibitem{strogatz} S. H. Strogatz, Nature (London) {\bf 410}, 268
(2001).

\bibitem{ws98} D. J. Watts and H. S. Strogatz, Nature {\bf 393}, 440
  (1998).

\bibitem{ama00} L. A. N. Amaral, A. Scala, M. Barth\'{e}l\'{e}mi, and
H. E. Stanley, {\em Proc. Nat. Acad. Sci.} {\bf 97}, 11149 (2000).

\bibitem{bar99} A.-L. Barab\'{a}si, and R. Albert, Science {\bf 286},
509 (1999); A.-L. Barab\'{a}si, R. Albert, and H. Jeong, Physica A
{\bf 272}, 173 (1999).

\bibitem{pv01a} R. Pastor-Satorras, and A. Vespignani,
Phys. Rev. Lett. {\bf 86}, 3200 (2001).

\bibitem{moreno02}
	Y. Moreno, R. Pastor-Satorras, and A. Vespignani,
	Eur. Phys. J. B {\bf 26}, 521 (2002).

\bibitem{virusreview} R. Pastor-Satorras, and A. Vespignani, {\em
Handbook of Graphs and Networks: From the Genome to the Internet},
eds. S. Bornholdt and H.G. Schuster (Wiley-VCH, Berlin, 2002).

\bibitem{av03} A. V\'{a}zquez, and Y. Moreno, Phys. Rev. E {\bf 67},
	015101(R) (2003).

\bibitem{n02b} M. E. J. Newman, Phys. Rev. E {\bf 66}, 016128 (2002).

\bibitem{havlin01} R. Cohen, K. Erez, D. ben-Avraham, and
S. Havlin, {\em Phys. Rev. Lett.} {\bf 85}, 4626 (2000).

\bibitem{newman00} D. S. Callaway, M. E. J. Newman, S. H. Strogatz,
and D. J. Watts, {\em Phys. Rev. Lett.} {\bf 85}, 5468 (2000).

\bibitem{bar00} R. Albert, H. Jeong, and A.-L. Barab\'{a}si,
Nature {\bf 406}, 378 (2000).

\bibitem{dave} Dave Kosiur, ``IP Multicasting: The Complete Guide to
Interactive Corporate Networks'', Wiley Computer Publishing, John
Wiley \& Sons, Inc, New York (1998).
 
\bibitem{deering} S.  Deering, ``Multicast routing in internetworks
and extended LANs'', in Proc. ACM Sigcom '88, page 55-64, Stanford,
CA, USA (1988).

\bibitem{vogels} Vogels, W., van Renesse, R. and Birman, K., "The
Power of Epidemics: Robust Communication for Large-Scale Distributed
Systems", in the Proceedings of HotNets-I, Princeton, NJ (2002).

\bibitem{ep3} A.-M. Kermarrec, A Ganesh and L. Massoulie,
  ''Probabilistic reliable dissemination in large-scale systems'', IEEE
  Trans. Parall. Distr. Syst., in press (2003).

\bibitem{p2p} {\em Peer-to-Peer: Harnessing the Power of Disruptive
Technologies}, ed. A. Oram (O'Reilly \& Associates, Inc., Sebastopol,
CA, 2001).

\bibitem{ep1} A. J. Demers, D. H. Greene, C. Hauser, W. Irish, and
J. Larson, {\em Epidemic Algorithms for Replicated Database
Maintenance}. In Proc. of the Sixth Annual ACM Symposium on Principles
of Distributed Computing, Vancouver, Canada, 1987.

\bibitem{grid} I. Foster and C. Kesselman, eds., {The Grid: Blueprint
for a Future Computing Infrastructure}, Morgan Kaufman, San Francisco
(1999).
 
\bibitem{zanette} D. H. Zanette, Phys. Rev. E {\bf 64}, 050901(R) (2001).

\bibitem{liu} Z. Liu, Y.-C. Lai, and N. Ye, Phys. Rev. E {\bf 67},
031911 (2003).

\bibitem{dkbook} D. J. Daley and J. Gani, {\em Epidemic Modeling}
(Cambridge University Press, Cambridge UK, 2000).

\bibitem{dk64} D. J. Daley, and D. G. Kendall, Nature {\bf 204}, 1118
  (1964).

\bibitem{pre98} J. B. G\'omez, Y. Moreno, and A. F. Pacheco,
Phys. Rev. E {\bf 58}, 1528 (1998).

\bibitem{jgr} Y. Moreno, A. M. Correig, J. B. G\'omez, and
A. F. Pacheco, J. Geophys. Res. {\bf B 106}, 6609 (2001).

\bibitem{mgp03} Y. Moreno, J. B. G\'{o}mez, and A. F. Pacheco,
Phys. Rev. E {\bf 68}, 035103(R) (2003).

\bibitem{mnv03} Y. Moreno, M. Nekovee, and A. Vespignani, {\em
cond-mat/0311212} (2003).

\bibitem{nm03} M. Nekovee and Y. Moreno, in preparation.

\bibitem{vw} A. Vazquez and M. Weigt, Phys. Rev. E {\bf 67}, 027101 (2003).

\bibitem{mv02} Y. Moreno, A. V\'{a}zquez, Europhy. Lett. {\bf
57}, 765 (2002).

\bibitem{note1} Note that for the $sr$ variant, the initial conditions
  should include the existence of at least one stifler. This has been
  taken into account by randomly selecting a node among all the
  remaining ignorants. Hence the initial conditions are now
  $I(0)=N-2$, $S(0)=1$, and $R(0)=1$. 

\end{thebibliography}
\end{document}